\documentclass[11pt]{article}
\usepackage{amsfonts}
\usepackage{times}
\usepackage{fullpage}

\newtheorem{definition}{Definition}

\newtheorem{claim}{Claim}

\newtheorem{lemma}{Lemma}

\newtheorem{theorem}{Theorem}

\newcommand{\remove}[1]{}
\newcommand{\qed}{\hfill\rule{2mm}{2mm}}
\newcommand{\mod}{\mbox{\rm mod}\:}

\newcommand{\prob}{\Pr}

\newcommand{\ket}[1]{|\,#1\rangle}

\newcommand{\bit}{\{0,1\}}
\newcommand{\ebit}{\{-1,+1\}}

\newcommand{\expct}{\mathbf{E}}
\newcommand{\E}{\mathbf{E}}

\newcommand{\zp}{\mathbb{Z}_p}
\newcommand{\bvec}[1]{\mathbf{#1}}

\newcommand{\sig}{\mbox{\sf SIG}}
\newcommand{\siggen}{\mbox{\sf sig\_gen}}
\newcommand{\sigsign}{\mbox{\sf sig\_sign}}
\newcommand{\sigver}{\mbox{\sf sig\_verify}}
\newcommand{\sk}{sk}
\newcommand{\vk}{vk}

\newcommand{\verify}{\mathsf{ver}}
\newcommand{\alg}{\mathcal{Z}}

\newcommand{\dom}{X}
\newcommand{\breaker}{\mathcal{B}}
\newcommand{\signoracle}{\mbox{\sf OSign}}
\newcommand{\oracle}{\mathcal{O}}
\newcommand{\funcc}{\mathcal{C}}
\newcommand{\sd}{\mathsf{SD}}

\newcommand{\sqsoracle}{\mathsf{SQS}}
\newcommand{\sqloracle}{\mathsf{SQL}}

\newenvironment{proof}{\begin{trivlist}
\item[\hspace{\labelsep}{\bf\noindent Proof: }]
}{\qed\end{trivlist}}
\newenvironment{proofof}[1]{\begin{trivlist}
\item[\hspace{\labelsep}{\bf\noindent Proof of #1: }]
}{\qed\end{trivlist}}
\newenvironment{sketch}{\begin{trivlist}
\item[\hspace{\labelsep}{\bf\noindent Proof sketch: }]
}{\qed\end{trivlist}}

\title{On Statistical Query Sampling and NMR Quantum Computing}
\author{Avrim Blum\thanks{
Computer Science Department, Carnegie Mellon
  University, 5000 Forbes Ave. Pittsburgh, PA 15213. E-mail: {\tt
    \{avrim,yangke\}@cs.cmu.edu}.  This work is supported in part by
NSF grants CCR-0105488 and NSF-ITR 0122581.}
\and
Ke Yang$^{*}$
}
 
\begin{document}
\maketitle

\begin{abstract}

We introduce a ``Statistical Query Sampling'' model, in which 
the goal of an algorithm is to produce an element in a hidden set
$S\subseteq\bit^n$ with reasonable probability. The algorithm
gains information about $S$ through oracle calls (statistical
queries), where the 
algorithm submits a query function $g(\cdot)$ and receives an
approximation to $\Pr_{x \in S}[g(x)=1]$.  We
show how this model is related to NMR quantum computing, in which only
statistical properties of an ensemble of quantum systems can be
measured, and in particular to the question of whether one can
translate standard quantum algorithms to the NMR setting without
putting all of their classical post-processing into the quantum
system.  Using Fourier analysis techniques developed in the related
context of {\em statistical query learning}, 
we prove a number of lower bounds (both information-theoretic and
cryptographic) 
on the ability of algorithms to produces an $x\in S$, even when the
set $S$ is fairly simple.  These lower bounds point out a difficulty
in efficiently applying NMR quantum computing to algorithms such as
Shor's and Simon's algorithm that involve significant classical
post-processing.  We also explicitly relate the notion of statistical
query sampling to that of statistical query learning.
\end{abstract}

\section{Introduction}

Recent years have witnessed the development of a number of exciting
quantum algorithms: Simon's algorithm for the hidden XOR secret
problem~\cite{Sim94}, Shor's algorithm for factoring and discrete
logarithms~\cite{Sho94,Sho97}, Boneh and Lipton's algorithm for the
hidden subgroup problem~\cite{BL95}, and many generalizations and
extensions~\cite{Joz97,EH99,EHK99,HRT00,GSVV01,H02}.  At the same
time, work has been ongoing on various proposals for physically
realizing quantum computers.  Currently, one of the most promising
such proposals is based on Nuclear Magnetic
Resonance (NMR)~\cite{DiV95,CFH97,GC97,CVZ+98}.  The NMR approach works by
manipulating a large ensemble of quantum systems in solution.  One
property of the NMR method, which is the focus of this paper, is that
unlike the ``standard'' quantum computing model, one cannot directly
measure any individual quantum system in the ensemble.  Instead, a
measurement is limited to a single qubit, and when
a measurement takes place, the device returns (an approximation to)
the expected value of this measurement, over the quantum systems in
the ensemble.  For this reason, the model for NMR is sometimes called
the ``expected-value'' (EV) model~\cite{Col02}. In contrast, the
measurement in 
the standard quantum model yields a random sample state (which may
consists of multiple bits) according to a classical probability
distribution. 

\remove{
More specifically, in the EV model, a quantum
system is initiated to state $\ket{0^n}$,\footnote{In fact,
initialization appears difficult to do well in NMR, and this
difficulty can lead to new problems~\cite{ASV00}, but that is not our
focus here.} and then passes through a
sequence of unitary operations --- just as in the standard
model.  However, the measurement of the quantum system in the EV model
is limited to a projective measurement of a single qubit, and the
result returned is an approximation to the expected value of this
measurement (or equivalently, the probability
that the measurement returns ``1'').  In contrast, the measurement in
the standard quantum model yields a random sample state (which can
consists of multiple qubits) according to a classical distribution.
}
Given the distinction between the standard model and the EV model, the
first question that arises is whether it is possible to 
translate algorithms working in the standard model to work in the
EV model. 
In fact, the answer is yes.  Consider any BQP algorithm~\cite{NC00}.
Recall from the definition that a BQP algorithm solves a decision
problem, and such an algorithm has a special ``target'' qubit to
indicate acceptance.  For a language $L$ and an input $x$, if $x\in
L$, then the measurement of the target qubit will produce a ``1'' with
probability at least $3/4$; if $x\not\in L$, the probability is at
most $1/4$ when measured. Such an 
algorithm works naturally in the EV model, since one can simply
measure the target qubit, and even with significant measurement error,
use the rule that if the observed value $v\ge1/2$, then $x\in L$, and
otherwise $x\not\in L$.  For a search (as 
opposed to decision) problem, we can perform the usual reduction to a
series of decision problems, solving each one by one. In fact, many
researchers have used this approach~\cite{GC97,NC00}, which we call an
``all-inclusive'' translation.

Unfortunately, the ``all-inclusive'' translation can greatly increase
the amount of work that must be done by the quantum system.  Consider
Shor's algorithm, for instance (see Appendix~\ref{app:simon-shor}).
Shor's algorithm (and others like it) 
consists of a quantum sampling circuit $Q$, whose output is measured
and fed into a classical extraction circuit $C$.  For the
all-inclusive translation, the classical extraction circuit $C$ needs
to be ``quantumized'', i.e., realized by a quantum circuit and
appended to the quantum sampling circuit $Q$.  This can cause a
significant increase in the size 
of the quantum circuit --- in the case of Shor's algorithm, the entire
circuitry for computing continued fractions needs to be realized in
quantum --- which is a rather undesirable consequence.  Even in the
most optimistic scenarios, quantum computers will be orders of
magnitude more difficult to manufacture and maintain than classical
computers, and thus we would like to put as little of the complexity as
possible in the quantum system.   Even more serious problems emerge
when 
more than one sample is needed by the classical extraction circuit.
For example, in Simon's algorithm, 
$\Omega(n)$ samples are needed for Gaussian elimination
(see Appendix~\ref{app:simon-shor}).  Now the all-inclusive
translation needs to manufacture multiple copies of the quantum
sampling circuit and then connect them together with the
``quantumized'' classical extraction circuit.  This can cause even
more blowup in the size of the quantum circuit in the EV model.


In this paper, we consider the question of whether there might be more
efficient translations that apply generally to algorithms consisting
of a quantum sampling circuit $Q$ followed by a classical extraction
circuit $C$, that work {\em without} having to put the classical part
of the algorithm into the quantum system.  Our main contributions are
results that answer this question in the negative, for several natural
notions of ``general''.  We achieve these results through a connection
to the notion of {\em statistical query learning} \cite{K93} studied
in Computational Learning Theory, and in particular to a related
notion that we introduce of {\em statistical query sampling}.  Using
techniques from Fourier analysis and cryptography, we show that even
in cases where the 
distribution implied by $Q$ is quite simple, it can be hard to use the
EV model to generate a sample that can be used by $C$.  
Note that our results do not preclude the possibility of approaches
tailored to specific quantum algorithms.  For example,
Collins~\cite{Col02} demonstrates a modification to Grover's algorithm
that is more efficient than the all-inclusive translation (see also
the discussion below).  However,
as pointed out by the author, his approach does not generalize
to algorithms like Shor's.

\subsection{Our model and results}
We view the quantum sampling circuit $Q$ as representing
a hidden set $S\subseteq\bit^n$, and we view the classical
post-processing as a circuit $C$ such
that $C(x) = 1$ for all $x \in S$.  The goal of the translation
procedure is to produce some $x \in S$.  To find such an $x$, the
algorithm has the ability to perform a ``statistical query'' of $Q$ by
proposing a query function (a predicate) $g:\bit^n\mapsto\bit$ and
asking for $\expct_{x \in S}[g(x)]$ up to some $1/poly$ accuracy.  For
example, measuring the $i$th qubit corresponds to the query $g(x) =
x_i$.   Taking the XOR of the first three qubits and then measuring the result
corresponds to the query $g(x) = x_1 \oplus x_2 \oplus x_3$.
The algorithm may repeat this process multiple
(polynomially-many) times, with different query functions $g$, and in
the end must (with noticeable probability) produce an $x\in S$.  

Note that this task is easy to do if $S$ is very large ($|S| \geq
2^n/poly(n)$), since a random $x \in \bit^n$ will do.   It is also
easy to do if $S$ is very small ($|S| = poly(n)$).  In particular, if $|S| =
poly(n)$, then by asking for an accuracy of $1/(2|S|)$ one can
distinguish the case that $\E_{x \in S}[g(x)] = 0$ from the case that
$\E_{x \in S}[g(x)]> 0$.  This allows one to walk down the bits of
$x$, fixing bits from left to right, until a specific $x\in S$ is
produced.  This is the key idea of \cite{Col02}.

We show, however, that this task is hard in general.  Specifically, we
give two types of hardness results.  First, we give 
an information-theoretic hardness result if the query algorithm is
not allowed to access $C$.  That is, the translator is allowed to use
the fact that the classical extraction circuit $C$ is polynomial in
size  (so the set of accepting strings cannot be totally arbitrary)
but it is not allowed to examine $C$ --- it can only gain
information via the queries $g$.  Second, we give a cryptographic
hardness result if we assume the translator is given $C$ as input,
but that otherwise $C$ is an arbitrary polynomial-size circuit.  We
still do not know if efficient translation is possible for the
specific circuit $C$ used in Shor's algorithm.

We also consider a more general setting, in which $S$ may be large
(e.g., $|S|= 2^{n-1}$), so a random string has reasonable chance of
belonging to $S$, but the goal of the translation is to produce a
string $x\in S$ with 
probability substantially greater than random guessing. We call this
more general setting ``strong SQ-sampling'', and refer to the former
setting as the ``weak SQ-sampling''. Strong SQ-sampling  models
situations such as Simon's algorithm, in which the quantum circuit
produces a random $y\in \bit^n$ such that $y\cdot s=0$ for the hidden
secret $s$.  In this case, a random string has probability $1/2$ of
belonging to $S$, but we need $\Omega(n)$ correct samples in a row in
order to perform Gaussian elimination. 
We give an information-theoretic hardness result for this
problem, that holds for the specific set $S$ used by Simon's
algorithm (Theorem
\ref{thm:parity-2}).\footnote{Note, for Simon's algorithm, we no
longer want to think of there existing a known classical extraction
circuit.  If we were given access to a circuit $C$ such that $C(x)=1$
iff $x \in S$ (e.g., the circuit with the hidden secret built in) then the
sampling goal would be easy. See
Theorem~\ref{thm:sq-sample-sq-learn} for further discussion.}

\subsection{Techniques and relation to Statistical Query learning}

Our results are based on a connection to the Statistical Query
(SQ) learning model, first introduced by Kearns~\cite{K93} as a
restricted version of the popular Probably Approximately Correct (PAC)
model of Valiant~\cite{V84}. In these learning models, the goal of
an algorithm is to learn an approximation to a hidden function $f:\bit^n
\mapsto \bit$.  In the PAC model, the algorithm has access to an
``example oracle'', which produces a random labeled sample $\langle
x,f(x)\rangle$ upon invocation.  In the SQ model, however, the
algorithm does not see explicit examples or their labels. Instead, the
algorithm queries an ``SQ-oracle'' with predicates $g(x,y)$, and
receives an approximation to $\prob_{x}[g(x,f(x))=1]$.  For instance,
the algorithm might ask for the probability that a random example
would both be positive and have its first bit set to 1 ($g(x,y) = x_1
\wedge y$).\footnote{In both PAC and SQ learning models, the
distribution over $x$ need not be the uniform distribution (or even
known to the learning algorithm).  However, much work on SQ learning
does focus on the uniform distribution, and that is the setting we are
most interested in in this paper.}  The SQ model has proven to be very
useful because (a) it is inherently tolerant to classification noise
(this is the reason the model was developed), and (b) nearly all
machine learning algorithms can be phrased as SQ algorithms. What
makes the SQ model especially interesting is that one can
information-theoretically prove lower bounds on the ability of SQ
algorithms to learn certain classes of
functions~\cite{K93,BFJ+94,J00,Yan01,Yan02}.

The relationship between the standard model and the EV model for
quantum 
computation is quite similar to that  between the PAC model and the SQ
model in machine learning, which motivates our definition of the
Statistical Query Sampling problem.  In particular, the SQ sampling
problem can be viewed as the SQ learning problem with two key
differences: first, the goal is not to learn an approximation to $f$
but is rather to produce a positive example, and second, the oracle for
SQ sampling 
returns approximations to $\prob[g(x)=1 \mid f(x)=1]$ rather than to
$\prob[g(x,f(x))=1]$ (a difference that matters when the set of
positive examples is quite small).

We use techniques from Fourier analysis to prove the
following lower bounds. First (Theorem~\ref{thm:bool-lin}) we show
there exist simple function classes such that no 
algorithm, using only a polynomial number of queries of $1/poly$
accuracy, can produce a positive instance with even $1/poly$
probability.  Second (Theorem~\ref{thm:parity-2}), for the class of
``negative parity'' functions arising in Simon's algorithm, no
algorithm using only a polynomial number of queries of $1/poly$
accuracy, can produce a nontrivial positive instance with probability
more than $1/2 + 1/poly$.   (Note that random guessing works with
probability $1/2$).  We also show
that unlike the case of SQ learning, the SQ sampling problem can be
computationally hard even if $f$ is explicitly given to the algorithm,
based on cryptographic assumptions
(see Theorem~\ref{thm:sig}). 

Finally, we explicitly relate the notion of SQ sampling to that of
SQ learning by proving that if a function class is ``dense'', meaning
that a random element has non-negligible probability of being positive,
then strong SQ-learnability implies strong SQ-samplability
(Theorem~\ref{thm:sq-sample-sq-learn}). We also point out that there
exists function classes that are perfectly SQ-samplable, yet not even
weakly SQ-learnable.

\section{Preliminaries and Definitions}
\label{sec:prelim}

We are interested in \emph{predicates} that map elements from a
\emph{domain} $\dom$ (e.g., $\bit^n$) to $\bit$. 
For a predicate $f:\dom\mapsto\bit$, 
an input $x$ is a \emph{positive input} to $f$ if $f(x)=1$, else it is
a \emph{negative input}. All the positive
inputs to $f$ form the \emph{positive set} of $f$, denoted by $S_f$.
A \emph{predicate class}, often denoted
by $\funcc_n$, is simply a collection of predicates over $\bit^n$.
A \emph{predicate class family} is an infinite sequence of predicate
classes $\funcc = (\funcc_1, \funcc_2, ...)$, such that
$\funcc_n$ is a predicate class over $\bit^n$.

A \emph{parity} function $\oplus_s(x)$ is defined to be 
$\oplus_s(x)=s\cdot x\;\mod\: 2$.
A \emph{negative parity} function $\lnot\oplus_s(x)$ is the negation
of the parity function $\oplus_s(x)$. 

\remove{
\begin{definition}[Density of Predicates]
\label{def:density-pred}
The \emph{density} of a predicate $f:\bit^n\mapsto\bit$ is the
fraction of its positive inputs
In other words, $\rho(f) = \prob_{x\in\bit^n}[f(x)=1]$.
%
A predicate class family $\funcc$ is \emph{dense} if there exists a
polynomial $p(\cdot)$ such that for every $n$ and  for every
$f\in\funcc_n$, $\rho(f)\ge 1/p(n)$.
A predicate class family $\funcc$ is \emph{sparse} if there exists a
negligible function $\kappa(\cdot)$ such that for every $n$ and  for
every $f\in\funcc_n$, $\rho(f)\le \kappa(n)$.
\end{definition}
}

\subsection{Statistical Query Sampling}

\begin{definition}[Statistical Query Sampling Oracle]
\label{def:sq-oracle-sample}
A \emph{statistical query sampling oracle} (SQS-oracle) for a
predicate $f$ is denoted by $\sqsoracle^f$. On input
$(g,\xi)$, where $g:\bit^n\mapsto\ebit$ is the \emph{query
  function} and $\xi\in[0,1]$ is the \emph{tolerance}, the oracle
returns a real   
number $y$ such that  $|y-\expct_{x\in S_f}[g(x)]|\le\xi$. 
\end{definition}

\begin{definition}[SQ-Samplability]

A predicate class family $\funcc$ is \emph{SQ-samplable at rate
  $s$ in time $t$ and tolerance $\xi$}, if there exists a randomized oracle
machine $\alg$, such that for every $n>0$ and every $f\in\funcc_n$,
$\alg$ with access to any SQS-oracle  $\sqsoracle^f$, runs in
at most $t(n)$ steps, asks queries with tolerance at least $\xi$, and
outputs an $x\in S_f$ with probability at least $s(n)$.
We say $\funcc$ is \emph{strong SQ-samplable} if 
for every $\epsilon$, $\funcc$ is SQ-samplable at rate $1-\epsilon$ in
time $t$ and tolerance $\xi$ such that $t$ and $\xi^{-1}$ are
polynomial in $n$ and $1/\epsilon$. 
We say $\funcc$ is \emph{weak SQ-samplable} if there exists a
polynomial $p$, such that $\funcc$ is SQ-samplable at rate $1/p(n)$ in
time and inverse tolerance polynomial in $n$.
\end{definition}

\remove{
\begin{definition}[Strong SQ-Samplable Function Class]
\label{def:strong-sqs-func}
A predicate class family $\funcc$ is \emph{strong SQ-samplable}, if
there exists a randomized oracle machines $\alg$, such that
for every $n>0$, every $f\in\funcc_n$ and for every $\epsilon>0$,
$\alg$ with access to an arbitrary SQS-oracle  $\sqsoracle^f$ outputs
an $x\in S_f$ with probability 
at least $1-\epsilon$, and furthermore, both the running time of
$\alg$ and the inverse of the tolerance of each query made by
it are bounded by a polynomial in $n$ and $1/\epsilon$.
\end{definition}

\begin{definition}[Weak SQ-Samplable Function Class]
\label{def:weak-sqs-func}
A predicate class family $\funcc$ is \emph{weak SQ-samplable}, if 
there exists a randomized oracle machines $\alg$ and a polynomial
$p(\cdot)$ such that for every $n>0$ and every $f\in\funcc_n$, $\alg$
with access 
to any SQS-oracle  $\sqsoracle^f$ outputs an $x\in S_f$ with
probability at least $1/p(n)$, and furthermore, both the running time
of $\alg$ and  the inverse of the tolerance of each query made by it
are bounded by a polynomial in $n$.
\end{definition}
}


\begin{definition}[Sampling Algorithms with Auxiliary Inputs]
\label{def:aux-input}
A predicate class family $\funcc$ is \emph{SQ-samplable
  with auxiliary input $\phi$} if it is SQ-samplable by an
algorithm $\alg$ which takes $\phi(f)$ as the auxiliary input, where
$f$ is the predicate being sampled.
\end{definition}

\section{Lower Bounds Based on Fourier Analysis}

We first prove two hardness results on SQ sampling, using Fourier analysis
techniques developed in the context of SQ learning. 

\remove{
We prove two hardness results on weak SQ-sampling. The first result is
an information theoretical one that proves the existence of a
very simple predicate class 
family that is not weak SQ-samplable.
The second result is
a cryptographic one, which demonstrates a predicate class family that
cannot be weak SQ-sampled even if the algorithm is given the full
description of the predicate, assuming the existence of one-way
functions.
}

\subsection{A Lower Bound on Weak SQ-Sampling}

We prove that there exist very simple families of
predicate classes that are not weak SQ-samplable, i.e., no
efficient algorithm can produce a positive input 
at any non-negligible rate.

We introduce a bit more notation.  We use boldface to
denote a vector and index the entries of an $n$-dimensional vector from
$0$ to $(n-1)$. We use $\bvec{x}[i]$ to denote the $i$-th entry of
$\bvec{x}$. $\bvec{x}[a..b]$ indicates the sub-vector formed by
the entries of $\bvec{x}$ between the $a$-th and the $b$-th,
inclusive.
Let $\hat{\dom}_{n,p}$ be the set of all $n$-dimensional
vectors over 
$\zp$ (the Galois field modulo $p$) whose last $n-1$ entries are not
all-zero, i.e., 
\begin{equation}
\label{eqn:dom-blin}
\hat{\dom}_{n,p}=\{\bvec{x}\in\zp^n\;|\;\bvec{x}[1..n-1]\ne
(0,0,...,0)\}.
\end{equation}
It is easy to see that $|\hat{\dom}_{n,p}|=p^n-p$.

\begin{definition}[Booleanized Linear Functions]
\label{def:bool-linear-p}
A \emph{booleanized linear function} over $\hat{\dom}_{n,p}$ with parameter $\bvec{a}$ is
denoted by $L_{\bvec{a}}$ and defined as
\begin{equation}
L_{\bvec{a}}(\bvec{x}) = \left\{
\begin{array}{lll}
1 & & \mbox{\rm if $\;\bvec{a}\cdot\bvec{x} = 1 \pmod p$}\\
0 & & \mbox{\rm otherwise}
\end{array}
\right.
\end{equation}
We say  $L_{\bvec{a}}$ is \emph{normalized} if
$\bvec{a}[0]=1$. 
The \emph{normalized booleanized linear function class}, denoted by
$\mathcal{L}_{n,p}$, consists of all normalized booleanized linear
functions over $\hat{\dom}_{n,p}$. In other words,
\begin{equation}
\label{eqn:dom_blin}
\mathcal{L}_{n,p}=\{L_{\bvec{a}}\;|\;\bvec{a}\in\zp^n,\; \bvec{a}[0]=1\}
\end{equation}
\end{definition}

\begin{theorem}
\label{thm:bool-lin}
If a sampling algorithm for the normalized booleanized linear function
class $\mathcal{L}_{n,p}$ makes less than $p^{n/4}$ queries, each of
tolerance $1/p^{n/3}$, then the probability it produces a
positive input $\bvec{x} \in \hat{\dom}_{n,p}$ is at most $1/p+1/p^{n/13}$.
\end{theorem}

Notice that the requirement $\bvec{x} \in \hat{\dom}_{n,p}$ is
simply to rule out the trivial positive input $100\ldots0$, and we
could have equivalently just modified the definition of a
``booleanized linear function'' so that this specific example is made
negative.  Also, notice that if we choose $p$ to be much greater than
$n$, say picking $p$ to be an $n$-bit prime number, then
$1/p+1/p^{n/13}$ is exponentially small, while the size of the problem
is still polynomial 
in $n$.  Furthermore, if a completely random $\bvec{x}$ is picked,
the probability it is a positive input is $1/p$. Thus even
exponentially many queries may only help the sampling by an
exponentially small margin.
 
\begin{proof}
Our proof strategy is similar to that used by Kearns~\cite{K93} and
Blum~et.~al.~\cite{BFJ+94} in the context of SQ learning.
We describe an ``adversarial'' SQS-oracle
$\widetilde{\sqsoracle}$ that
does not commit to any particular predicate at the beginning.
Rather, the oracle maintains a ``candidate predicate set''
$P$, which initially includes all predicates in 
the class $\mathcal{L}_{n,p}$ (a total $p^{n-1}$ of them). Each time
the algorithm $\alg$ makes a query, $\widetilde{\sqsoracle}$ replies
with an answer 
that yields very little information. Some predicates in the candidate
set $P$ might not be consistent with the answer and will be removed
from set $P$. After all the queries are finished,
$\widetilde{\sqsoracle}$ then 
commits to a random predicate remaining in $P$. We shall prove that
each query only removes a small fraction of the predicates from
$P$. Thus if $\alg$ does not make enough number of queries, there
would be enough predicates left in $P$ such that no element can be
positive with high probability.

For a query function $g:\hat{\dom}_{n,p}\mapsto\ebit$, we say that a
subset 
$S\subseteq\bit^n$ is a \emph{$\xi$-independent subset} for $g$, if 
$|\expct_{x\in S}[g(x)] - \expct_{x\in\hat{\dom}_{n,p}}[g(x)]|\le\xi$,
and we 
say a predicate $f$ is \emph{$\xi$-independent} from $g$, if its
positive set $S_f$ is a $\xi$-independent set for $g$. Intuitively, if
a predicate $f$ is $\xi$-independent from $g$, then the query
$(g,\xi)$ reveals almost no information about $f$, since
$\widetilde{\sqsoracle}$  can reply with 
$\expct_{x\in\hat{\dom}_{n,p}}[g(x)]$
instead, which is completely independent from $f$.

We describe the behavior of our SQS-oracle $\widetilde{\sqsoracle}$ in
more detail.
On query $g$, $\widetilde{\sqsoracle}$ replies
with  $\expct_{x\in\hat{\dom}_{n,p}}[g(x)]$, and removes all predicates
that are 
not $\xi$-independent from $g$ from the candidate set $P$. We
assume that all queries have tolerance $\xi= p^{-n/3}$. We shall
prove that for any query $g$, there are at most $p^{2n/3+2}$
predicates 
not  $p^{-n/3}$-independent from $g$. This proof is by a
Fourier analysis technique and is given as
Lemma~\ref{lemma:bias-fourier} in Appendix~\ref{app:proof}. Thus,
if less than $p^{n/4}$ queries are made, the candidate set still
contains at least $p^{n-1}(1-p^{-n/12-3})$ parity functions. 

Now consider the domain $\hat{\dom}_{n,p}$. It is not
hard to see that every $x\in\hat{\dom}_{n,p}$ is positive 
for only $p^{n-2}$ predicates.  So, if the oracle commits to a random
predicate out of the set of $p^{n-1}(1-p^{-n/12-3})$, the probability
that $x$ 
is positive is at most $1/p+1/p^{n/13}$. 
\end{proof}

\subsection{A Lower Bound on Sampling Negative Parity Predicates}

We prove that a class of negative  parity functions is not
SQ-samplable in polynomial time at any rate non-negligibly higher than
$1/2$.  


\begin{theorem}
\label{thm:parity-2}
Let $\dom_n = \bit^n\backslash\{0^n\}$ and $\funcc_n$ be the class of
negative parity functions over $\dom_n$. If a sampling algorithm
for $\funcc_n$ makes less than $2^{n/4}$ queries, each of
tolerance $2^{-n/4}$, then 
the probability it produces a positive input is at most
${1\over2}+{1\over2^{n/4-2}}$.
\end{theorem}

Before proving the theorem, we point out how this result relates to
the translation of Simon's algorithm to the NMR model. In Simon's algorithm,
the quantum sampling circuit produces a random $y\in \bit^n$ such that
$y\cdot s=0$, where $s$ is the ``hidden'' secret (see
Appendix~\ref{app:simon-shor}). Thus the hidden set corresponds
exactly to
the negative parity function $\lnot\oplus_s$. In the algorithm, the
quantum sampling circuit is invoked $\Theta(n)$ times and produces
$\Theta(n)$ samples for Gaussian elimination. Notice that $y=0^n$ is
useless. Therefore, a translation of the quantum
sampling circuit will produce an SQ-sampling algorithm $\alg$ to be
executed $\Theta(n)$ times and to produce $\Theta(n)$ positive samples
in $\dom_n = \bit^n\backslash\{0^n\}$. However,
Theorem~\ref{thm:parity-2} implies that it is not possible to
sample efficiently at any rate non-negligibly higher than $1/2$ 
(notice that a random $x\in\dom_n$ is positive with probability
almost $1/2$). This result suggests that it appears necessary to
manufacture 
$\Theta(n)$ copies of the quantum sampling circuit and run these copies
together in the NMR model.

\remove{
suggests that it
unlikely to efficiently translate the quantum sampling circuit in
Simon's algorithm to the EV model. In Simon's algorithm, the quantum
sampling circuit produces a random $y\in \bit^n$ such that 
$y\cdot s=0$, where $s$ is the ``hidden'' parameter (see
Appendix~\ref{app:simon-shor}). Thus the hidden set corresponds exactly
the negative parity function $\lnot\oplus_s$. In the algorithm, the
quantum sampling circuit is invoked $\Omega(n)$ times and produces
$\Omega(n)$ samples for Gaussian elimination. Notice that $y=0^n$ is
useless for Gaussian elimination.
Therefore, a generic translation of the quantum
sampling circuit will produce an SQ-sampling algorithm $\alg$ that 
will be executed $\Omega(n)$ times to produce samples in 
$\dom_n = \bit^n\backslash\{0^n\}$. However, by
Theorem~\ref{thm:parity-2}, 

If the
probability that such an algorithm produces a positive input is less
than $2/3$, then a constant fraction of the samples produced by $\alg$
will be ``wrong'', and the Gaussian elimination would not work. In fact,
for the generic translation to have even a constant success
probability, the sampling
algorithm needs to produce a positive sample with probability at least 
$1-O({1\over n})$.
}

\begin{sketch}
The proof strategy is similar to that of
Theorem~\ref{thm:bool-lin}. We assume that each query has tolerance
$\xi = 1/2^{n/4}$. We construct an SQS-oracle that on query function
$g$, replies
with  $\expct_{x\in\bit^n}[g(x)]$, and remove all predicates that are
not $\xi$-independent from $g$ from the candidate set $P$ (here the
definition of ``$\xi$-independent'' naturally changes to
$|\expct_{x\in S}[g(x)]-\expct_{x\in\bit^n}[g(x)]|\le\xi$). We shall
prove in Lemma~\ref{lemma:independent-count} (in
Appendix~\ref{app:proof}) that for any query $g$, there are at most
$2^{n/2+2}$ predicates 
not  $2^{-n/4}$-independent from $g$.  Thus,
if less than $2^{n/4}$ queries are made, the candidate set still
contains at least $2^n-2^{{3n/4}+2}-1$ parity functions. 

Now consider the domain $\dom_n=\bit^n\backslash\{0^n\}$. It is not
hard to see that every $x\in\dom_n$ is positive for 
$2^{n-1}$ negative parity functions. Now if a random parity function
is chosen from a set of size $2^n-2^{3n/4+2}-1$, the probability that
$x$ is positive is at most 
$${2^{n-1}\over 2^n-2^{3n/4+2}-1}\le {1\over2}+{1\over2^{n/4-2}}.$$
This is true for any $x\in\dom_n$. Therefore, whatever $\alg$ outputs,
the probability that it is 
positive is at most ${1\over2}+{1\over2^{n/4-2}}$.
\end{sketch}

\section{A Cryptographic Lower Bound}

We next prove a cryptographic lower bound.  Assuming that
one-way functions exist, we show that there exist predicate class
families 
that are not weak SQ-samplable, even if the sampling algorithm is
given the complete description of the predicate as the auxiliary
input. The technique we use here is somewhat similar to that of
Angluin and Kharitonov~\cite{AK91}, who used signature schemes to
prove that membership queries do not help to learn DNF.

We briefly describe the ideas behind our proof. We will use a
digital signature scheme secure against adaptive chosen
message  attack~\cite{GMR88}, which exists if one-way functions
exist~\cite{R90}. 
Let the predicate be the signature verification function
$\verify_{\vk}(m,s)$, which returns $1$ if $s$ is a valid signature to
message $m$ with respect to the verification key $\vk$.  The security
of the signature scheme states that no ``breaker'' $\breaker$, given
access to a signing oracle, can produce a {\em new} valid signature it has
not yet seen.  We want to argue that this implies no sampling
algorithm $\alg$, given access to a SQ-sampling oracle, can produce
{\em any} valid signature.  We will show that if such an algorithm
$\alg$ exists, we can construct a ``breaker'' $\breaker$ as follows.
The breaker will have access to a signing oracle $\signoracle$ that
signs any message given to it as input, and runs $\alg$ as a
subroutine.  The only non-trivial part for $\breaker$ is to simulate
an SQS-oracle used by $\alg$ without revealing to $\alg$ any
information about which signatures it has already seen (so that $\alg$
is not biased towards producing an already-seen signature).  Upon a
query $(g,\xi)$ from $\alg$, $\breaker$ will produce a number of
random messages, ask the signing oracle to sign them, and use these
samples to estimate $\expct_{x\in S_f}[g(x)]$. Next, $\breaker$
``randomizes'' this estimate by adding an artificial noise to it. With
properly chosen parameters, this ``randomized'' estimate is still
a valid answer with very high probability, and yet almost independent
from the messages $\breaker$ produces. Finally, $\alg$ produces a
positive input, which is a message/signature pair $(m',s')$. The
distribution of the this pair $(m',s')$ is also almost independent
from the messages $\breaker$ produces, and if $\alg$ only makes
polynomially many queries, then only polynomially many messages will
be produced by $\breaker$.  Therefore the probability that $m'$ is one
of the messages produced by $\breaker$ is very small, and so
$\breaker$ breaks the digital signature scheme with reasonably high
probability. 

Formally, a
\emph{signature   scheme} \sig\ 
is a triple $(\siggen,\sigsign,\sigver)$ of algorithms, the first two
being probabilistic, and all running in polynomial time.  $\siggen$
takes as input $1^n$ and outputs a signing/verification key pair
$(\sk,\vk)$.  \sigsign\ takes a message $m$ and a signing key $\sk$ as
input and outputs a signature $s$ for $m$.  WLOG we assume that both
$m$ and $s$ are $n$-bits long. \sigver\ takes a message $m$, a
verification key $\vk$, and a candidate signature $s'$ for $m$ as
input and returns the bit $b=1$ if $s'$ is a valid signature for $m$
for the corresponding verification key $\vk$, and otherwise returns
the bit $b=0$.  Naturally, if $s= \sigsign(\sk,m)$, then
$\sigver(\vk,m,s) = 1$.  In an adaptive chosen message
attack~\cite{GMR88}, an adversary (``breaker'') $\breaker$ is given
$\vk$, where 
$(\sk,\vk)\leftarrow \siggen(1^n)$, and tries to forge signatures with
respect to $\vk$.  The breaker $\breaker$ is allowed to query a
signing oracle $\signoracle_{\vk}$, which signs any message with
respect to $\vk$, on messages of its choice. It succeeds in
existential forgery if after 
this it can output a pair $(m,s)$, where $\sigver(\vk,m,s)=1$, but $m$
was not one 
of the messages signed by the signature oracle.
A signature scheme $\sig$ is existentially unforgeable against
adaptive chosen message attacks if there is no forging algorithm
$\breaker$ that
runs in time polynomial in $n$ and succeeds with probability
$1/poly(n)$.  Such schemes exist if one-way functions exist \cite{R90}.


\begin{theorem}
\label{thm:sig}
Let $\sig=(\siggen,\sigsign,\sigver)$ be a digital signature scheme
secure against adaptive chosen message attack. Then the predicate
class family $\funcc_{n}=\{\verify_{\vk}\}$ is not weakly
SQ-samplable, even if the sampling algorithm is given $\vk$ as the
auxiliary input. Here $\verify_{\vk}$ is defined to be
$\verify_{\vk}(m,s)=\sigver(\vk,m,s)$, where
$(\sk,\vk)\gets\siggen(1^n)$, and $m,s\in\bit^n$.  
\end{theorem}

\begin{proof}
Assume to the contrary that there exists an algorithm $\alg$ that weak
SQ-samples the function class $\funcc_{n}=\{\verify_{\vk}\}$. More
precisely, we assume that $\alg$ produces a positive input with
probability $\epsilon$ by making $q$ queries, where both $1/\epsilon$
and $q$ are bounded by a polynomial in $n$.
We shall construct a polynomial-time algorithm $\breaker$ that breaks
the signature scheme $\sig$ with probability $\epsilon/2$, causing a 
contradiction. 

We now describe the behavior of  $\breaker$. $\breaker$ has access to a
signing oracle $\signoracle_{\vk}$ and interacts with the sampling
algorithm $\alg$ as the SQS-oracle. When $\alg$ makes a query
$(g,\xi)$, $\breaker$ does the following. First, $\breaker$ computes 
$\xi_0 = {\xi\cdot \epsilon\over10 q}$ and 
$M={2\ln(10 q/\epsilon)\over\xi_0^2}$. Then $\breaker$ draws $M$
random messages $m_1, m_2, ..., m_M\in\bit^n$, and asks the signing
oracle to sign all of them. Assume the signatures are 
$s_1, s_2, ..., s_M$. Next, $\breaker$ uses these message/signature
pairs to estimate the expected value of $g$ by computing 
$x = {1\over M}\sum_{k=1}^Mg(m_k,s_k)$. Then $\breaker$ ``randomizes''
$x$ by drawing a $y$ uniformly randomly from the interval 
$[x-{\xi\over2}, x+{\xi\over2}]$,
and sending $y$ to $\alg$ as
the answer to the query $(g,\xi)$. $\breaker$ also maintains a
``history set'' set $H$ of all the messages it has generated, which is
initially $\emptyset$. After a query from $\alg$ is
answered, $\breaker$ adds the messages $m_1, m_2, ..., m_M$ to set
$H$.

After all the $q$ queries are made, $\alg$ produces a pair
$(m',s')$. If $\verify_{\vk}(m',s')=1$ and $m'\not\in H$, then
$\breaker$ outputs $(m',s')$ and successfully forges a
signature. Otherwise $\breaker$ aborts and announces failure.

It is clear that $\breaker$ runs in polynomial time.
Intuitively, we can show that after the randomization, with high
probability the sample $(m',s')$ produced by $\alg$ is almost
independent from the history set $H$. Therefore, with high
probability, $m'\not\in H$, and so $\breaker$ will succeed. More
precisely, we prove that $\breaker$ will succeed with probabilit at
least $\epsilon/2$.

We use $S_{\vk}$ to denote the positive set for predicate
$\verify_{\vk}$. In other words, $S_{\vk}$ consists of valid
message/signature pairs with respect to the verification key $\vk$.

\begin{claim}
\label{claim:sample}
For a query function $g$, if we define 
$\sigma= \expct_{(m,s)\in  S_{\vk}}[g(m,s)]$, then 
with probability at least $1-\epsilon/5q$, we have 
$|x-\sigma|\le\xi_0$ (all quantities are as defined in the proof
sketch of Theorem~\ref{thm:sig}). 
\end{claim}
\begin{proof}
This is due to a straightforward application of the Hoeffding
Bound. Each sample $(m_k,s_k)$ is an 
independent random element from $S_{\vk}$ and thus
$\expct_{(m,s)\in S_{\vk}}[g(m,s)=1]=\sigma$. So the expected value of
$x$ is $\sigma$. Now, the probability that $M$ independent samples 
yields an average below $\sigma - \xi_0$ is at most
$e^{-M\xi_0^2/2}$ (notice that the range of $g$ is $\ebit$). 
 Also the probability that the average is above
$\sigma+\xi_0$ is at most $e^{-M\xi_0^2/2}$. 
Therefore with probability at least 
$1-2e^{-M\xi_0^2/2}\ge 1-\epsilon/5q$, we   have 
$|x-\sigma|\le \xi_0$. 
\end{proof}

We fix a set consisting of $M$ message/signature pairs generated by
$\breaker$ in response to a query $(g,\xi)$, and denote this by $U$: 
$U=\{(m_k,s_k)\}_{k=1}^M$. We call this set a \emph{sample set}.
We say $U$ is \emph{typical}, if the
average $g(m_k,s_k)$ is indeed $\xi_0$-close to $\sigma$. By
Claim~\ref{claim:sample}, at most $\epsilon /5q$ fraction of the
sample sets 
are not typical. 

Notice that a typical sample set will yield an average that is
$\xi_0$-close to $\sigma$. This is a much higher accuracy than
required by the $\alg$, which has a tolerance of
$\xi$. However, $\breaker$ needs this accuracy to perform the
randomization. 
\begin{claim}
\label{claim:typical-safe}
If $U$ is a typical set, then the answer from $\breaker$ for this
query is valid.
\end{claim}
\begin{proof}
Notice that if $U$ is typical, then the average $x$ is $\xi_0$-close
to the true value $\sigma$. After the randomization, it is 
$(\xi_0+\xi/2)$-close to $\sigma$. This is less than $\xi$.
\end{proof}

We consider the distribution of the answer produced by $\breaker$ for
a particular query $(g,\xi)$. We denote this distribution by $D_U$,
where $U$ is the sample set used by $\breaker$.
\begin{claim}
If both $U_0$ and $U_1$ are typical sets, then the statistical
distance between $D_{U_0}$ and $D_{U_1}$ is at most $\epsilon/5q$.
\end{claim}
\begin{proof}
We use $x_0$ and $x_1$ to denote the averages obtained from $U_0$ and
$U_1$, respectively. If both $U_0$ and $U_1$ are typical, we have
$|x_0-\sigma|\le\xi_0$ and $|x_1-\sigma|\le\xi_0$. Thus we have
$|x_0-x_1|\le2\xi_0$. Notice that $D_{U_0}$ is a uniform distribution 
over the interval of length $\xi$ centered at $x_0$, and $D_{U_1}$ a
uniform distribution of same length centered at $x_1$. The claim
follows from Lemma~\ref{lemma:uniform-sd}.
\end{proof}

Notice the history set $H$ consists of $q$ sample
sets. We say a history set $H$ is \emph{typical}, if all its sample
sets are typical. Then at most $\epsilon/5$ fraction of the history
sets are 
not typical. We denote the distribution of all answers produced
by $\breaker$  using history set $H$ by $T_H$. 

\begin{claim}
If both
$H_0$ and $H_1$ are typical, then the statistical distance between
$T_{H_0}$ and $T_{H_1}$ is at most $\epsilon/5$.
\end{claim}
\begin{proof}
This directly follow the sub-additivity of statistical distance (see
Appendix~\ref{app:sd}). 
\end{proof}

Now we fix an arbitrary typical set $\tilde{H}$ and denote its
corresponding distribution of the answers by $\tilde{T}$. Then we know
the distribution from any typical set is at most $\epsilon/5$ away
from $\tilde{T}$.  

The only information $\alg$ receives from $\breaker$ is
represented by the distribution of the answers produced by $\breaker$,
which is in turn determined by the history set $\breaker$ uses. Thus,
the distribution of the pair $(m',s')$ is completely determined
by the history set $H$, and we denote this distribution by $O_H$. We
know that if $H$ is typical, then 
$\prob_{(m,s)\in O_H}[\verify_{\vk}(m,s)=1]\ge \epsilon.$
We fix the distribution $\tilde{O}$ that
corresponds to the history set $\tilde{H}$.  Then we have
\begin{equation}
\label{eqn:prob-succ}
\prob_{(m,s)\in \tilde{O}}[\verify_{\vk}(m,s)=1]\ge \epsilon.
\end{equation}
Furthermore, we know that for any
typical history set $H$, its corresponding distribution of $O_H$ is 
$\epsilon/5$-close to $\tilde{O}$.

Consider a new experiment (a new execution of the breaker $\breaker$)
that is identical to the original one, except 
when $\alg$ outputs a pair $(m',s')$, it does so according to the fixed
distribution $\tilde{O}$. 
\begin{claim}
Let  $\hat{M}$ be the maximum size of the sample sets in $\tilde{H}$.
Then the  probability of the new experiment is 
at least $\epsilon-\hat{M}\cdot q/2^n$.
\end{claim}
\begin{proof}
Notice that  the output of $\alg$ is independent from the history set
$H$. Moreover, the history set contains at most $\hat{M}\cdot q$
messages. So the probability that a particular $m$ is in $H$ is at
most  $\hat{M}\cdot q/2^n$. This fact, along with
(\ref{eqn:prob-succ}), proves the claim.
\end{proof}

Now
putting things together, with probability at most $\epsilon/5$, the
history set $H$ is not typical; if $H$ is typical, the difference
between the probabilities of the two experiments is at most
$\epsilon/5$; the probability of success of the new 
experiment is at least $\epsilon-\hat{M}\cdot q/2^n$. Therefore the
probability of success of the original experiment is at least (for $n$
large enough)
$\epsilon-\hat{M}\cdot q/2^n - \epsilon/5-\epsilon/5>\epsilon/2$.

This finishes the proof.
\end{proof}

\section{SQ sampling and SQ learning}

We now point out relationships between our SQ sampling model and the
SQ learning model of Kearns~\cite{K93}.  We begin with definitions of
SQ learning.  (In these definitions, we assume learning is with
respect to the uniform distribution over examples.)

\begin{definition}[Statistical Query Learning Oracle]
\label{def:sq-oracle-learn}
A \emph{statistical query learning oracle} (SQL-oracle) for a
predicate $f$ is denoted by $\sqloracle^f$. On an input
$(g,\xi)$, where $g:\bit^n\times\bit\mapsto\ebit$ is the
\emph{query function} and $\xi\in[0,1]$ is the \emph{tolerance}, the
oracle returns a real  
number $y$ such that  $|y-\expct_{x\in \bit^n}[g(x,f(x))]|\le\xi$. 
\end{definition}

\begin{definition}[Strong SQ-Learnability]
\label{def:sql-func}
A predicate class family $\funcc$ is \emph{Strong SQ-learnable} if 
there exists a randomized oracle machine $\alg$, such that
for every $n>0$, every $f\in\funcc_n$ and for every $\epsilon>0$,
$\delta>0$, $\alg$ with access to any SQL-oracle
$\sqloracle^f$ outputs a hypothesis 
$\hat{f}$ such that $\prob_{x\in\bit^n}[\hat{f}(x)=f(x)]\ge 1-\epsilon$
with probability at least $1-\delta$, and furthermore, both the
running time of $\alg$ and the inverse of the tolerance of each
query made by it are bounded by a polynomial in $n$,
$1/\epsilon$ and $1/\delta$. Here $\epsilon$ is called the
\emph{accuracy} and $\delta$ the \emph{confidence}.
\end{definition}

\begin{definition}[Weak SQ-Learnability]
\label{def:weak-sql-func}
A predicate class family $\funcc$ is \emph{weak SQ-learnable} if 
there exists a randomized oracle machines $\alg$ and a polynomial
$p(\cdot)$, such that for every $n$ and for every $f\in\funcc_n$,
$\alg$ with access to any SQL-oracle  $\sqloracle^f$, outputs a
hypothesis $\hat{f}$ such that 
$\prob_{x\in\bit^n}[\hat{f}(x)=f(x)]\ge 1/2+1/p(n)$,
and furthermore, both the running
time of $\alg$ and the inverse of the tolerance of each query made by
$\alg$ are bounded by a polynomial in $n$.
\end{definition}

The first observation to make is that a predicate class can be
strongly SQ-learnable and yet not even weakly SQ-samplable.  In particular,
any class with a sufficiently low density of positive examples can be
trivially learned by producing the ``all zero'' hypothesis.
(Formally, if we wish be correct even for values of $\epsilon$ that are
exponentially small, it suffices to have the density less than
$1/2^{n/2}$ so that if necessary we can use the SQL oracle to
identify all positive examples.)  In the other direction, a class can
be strongly SQ-samplable and yet not even weakly SQ-learnable.
Indeed, the family of negative parity functions taken over the domain
$\bit^n$ is trivially SQ-samplable (because $f(0^n)=1$ for any such
$f$), but such functions are not even weakly
SQ-learnable~\cite{K93}.  It is interesting to compare this to
Theorem~\ref{thm:parity-2}, since the predicate class families 
in these two theorems are very similar (one can think of the
difference either as removing $0^n$ from the domain, or simply as
changing the values of the functions at this one point),
yet they have completely different characterization in terms of
SQ-samplability.  

However, we show there is a relationship between these notions when
the set of positive examples is sufficiently dense.

\remove{
First, a learning
algorithm needs to produce a hypothesis that predicts the
labels of a random sample well, while a sampling algorithm only needs
to product a single positive input. Second, the scope of the
probability answer by an SQS-oracle is the positive set of the
predicate, while the scope of the probability by an SQL-oracle is the
entire domain.  
}

\subsection{SQ-learnability sometimes implies SQ-samplability}

We prove that under certain circumstances, SQ-learnability implies
SQ-samplability.

\begin{definition}[Density of Predicates]
\label{def:density-pred}
The \emph{density} of a predicate $f:\bit^n\mapsto\bit$, denoted by
$\rho(f)$,  is the fraction of its inputs that are positive.
In other words, $\rho(f) = \prob_{x\in\bit^n}[f(x)=1]$.
\end{definition}

\begin{definition}[Dense Predicates]
\label{def:dense-pred}
A predicate class family $\funcc$ is \emph{dense} if there exists a
polynomial $p(\cdot)$ such that for every $n$ and  for every
$f\in\funcc_n$, $\rho(f)\ge 1/p(n)$.
\end{definition}

\begin{theorem}
\label{thm:sq-sample-sq-learn}
If a dense predicate class family is strong SQ-learnable, then it is
also strong SQ-samplable with the auxiliary input $\rho$.
\end{theorem}

\begin{proof}
Let  $\alg$ be the algorithm that strongly SQ-learns dense predicate
family $\funcc$. 
We construct a new algorithm $A$ that strong SQ-samples $\funcc$ using
the density $\rho$ of the predicate $f$ as auxiliary input.  $A$ runs
a copy of $\alg$, whose accuracy and confidence are set to be
$\epsilon=\rho\cdot\epsilon'/4\ln({4\over\epsilon'})$ and
$\delta=\epsilon'/4$,  and simulates the
SQL-oracle used by $\alg$. We shall prove that $A$ produces a positive
input with probability at least $1-\epsilon'$.

We now describe the behavior of $A$. $A$ works in
two phases. In this first phase, it simulates the 
SQL-oracle $\sqloracle^f$.  When $\alg$ submits a query
$(g,\xi)$ to $A$, $A$ does the following.
\begin{enumerate}
\item
Set $M={9\ln(2q/\delta)\over2\xi^2}$, draw $M$ independent samples 
$x_1, x_2, ...,x_M$ from $\bit^n$, and compute 
$$s = {1\over M}\sum_{i=1}^Mg(x_i,0).$$
\item
Construct two query functions
$g_0(x)=g(x,0)$ and $g_1(x)=g(x,1)$. Submit queries
$(g_0,\xi/3)$ and $(g_1,\xi/3)$ to the
SQS-oracle $\sqsoracle^f$ and receive $y_0$ and $y_1$ as
answers. 
\item 
Compute $y = s +(y_1-y_0)\cdot \rho$ and send $y$ to $\alg$ as the
answer to the query $(g,\xi)$.
\end{enumerate}
The algorithm $A$ enters the second phase when $\alg$ produces a
hypothesis $\hat{f}$. Then $A$ repeats the following procedure. It
draws a random $x\in \bit^n$, and check if $\hat{f}(x)=1$. If so it
stops and output $x$; otherwise it continues. The procedure is
repeated $\ln\left({1\over\delta}\right)/\rho$ times and if $A$ still
hasn't stopped, it 
produces a random $x\in\bit^n$ and outputs it.

It is clear that $A$ runs in polynomial time. Now, we prove that $A$
produces a positive sample with 
probability at least $1-\epsilon'$.

First, we prove that with probability at least $1-\delta$, all answers
provided by $A$ are valid in the first phase.
Consider an average $s$ as an approximation of 
$\expct_{x\in\bit^n}[g(x,0)]$. We say $s$ is ``bad'', if 
$|s-\expct_{x\in\bit^n}[g(x,0)]|>\xi/3$. Then a simple application of
the Hoeffding Bound (see Appendix~\ref{app:hoeffding}) proves that
the probability that $s$ is bad is at most $\delta/q$.

Next, notice that 
$$g(x,f(x)) = g(x,0) + \left[g(x,1) - g(x,0)\right]\cdot f(x).$$
Therefore we have
\begin{eqnarray*}
\hspace{-0.1in}
\expct_{x\in\bit^n}[g(x,f(x))] 
& = &
\expct_{x\in\bit^n}[g(x,0)] + \\
& & 
\expct_{x\in\bit^n}[(g(x,1)-g(x,0))\cdot f(x)]\\
& = & \expct_{x\in\bit^n}[g(x,0)] + \\
& & \left(\expct_{x\in S_f}[g(x,1)] -
\expct_{x\in S_f}[g(x,0)]\right)\cdot \rho
\end{eqnarray*}

Therefore, if $s$ is not bad, then the $y$ computed by $A$ is a valid
reply to query $(g,\xi)$. Since $\alg$ makes a total of $q$ queries,
with probability at least $1-\delta$, all the replies by $A$ are valid
and $\alg$ should perform well. 

Next, consider the second phase of $A$. With probability at
least $1-\delta$, $\alg$ should produce a hypothesis $\hat{f}$ that
agrees with $f$ with probability at least $1-\epsilon$. Let us assume
th $\alg$ does produce such a $\hat{f}$. Now since a $\rho$ fraction
of the inputs are positive, the probability that 
$A$ doesn't draw a positive input in
$\ln\left({1\over\delta}\right)/\rho$  rounds is at most
$\delta$. The probability that $\hat{f}$ makes a mistake in any of the
rounds is at most 
$\ln\left({1\over\delta}\right)\cdot \epsilon/\rho$. If
$\hat{f}$ doesn't make any mistakes and at least one positive input
is drawn, then $A$ will correctly output it.

Putting everything together, we know that with probability at least
$1-3\delta-\ln\left({1\over\delta}\right)\cdot\epsilon/\rho=1-\epsilon'$,
$A$ 
will output a positive input.
\end{proof}

We remark that it appears necessary for the SQ-sampling algorithm to
have the density $\rho$ as an auxiliary input. One difference between
SQ-sampling and the SQ-learning is the \emph{resolution}. In the reply
of an SQS-oracle, the underline distribution is uniform over the
``hidden set'' $S_f$; for an SQL-oracle, the distirbution is uniform
over the entire set $\bit^n$. Therefore, a sampling algorithm needs
to know the size of $S_f$ in order to perform the simulation (more
precisely, in step 3 of the first phase).

It is interesting to compare this result to Theorem~\ref{thm:sig},
which shows a predicate class family that is perfectly SQ-learnable,
but not even weakly SQ-samplable. Nevertheless, there is no
contradiction  
since the predicate class family in Theorem~\ref{thm:sig} is not
dense. 

\remove{
\subsection{SQ-samplability Does Not Imply SQ-learnability }

We show that there exist families of predicate classes that are
strong SQ-samplable but not even weak SQ-learnable. 

\begin{theorem}
\label{thm:parity-1}
Let $\funcc$ be the family of the negative parity functions over
$\bit^n$.
Then $\funcc$
is strongly SQ-samplable but not weakly SQ-learnable.
\end{theorem}

\begin{proof}
It is easy to see that $f(0^n)=1$ for any $f\in\funcc_n$, and thus the
algorithm that always outputs $0^n$ produces a positive input with
probability 1. On the other hand, the fact that the class of parity
functions is not weak SQ-learnable was proven by Kearns~\cite{K93}.
\end{proof}

It is interesting to compare Theorem~\ref{thm:parity-1} to
Theorem~\ref{thm:parity-2}, since the predicate class families
in these two theorems are very similar, yet they have completely
different SQ-samplability. 
}


\section*{Acknowledgements}
We would like to thank David Collins for bringing this problem to our 
attention, and Bob Griffiths and David Collins for helpful discussions.

\appendix

\section{Shor's Algorithm and Simon's Algorithm}
\label{app:simon-shor} 

We briefly summarize Shor's algorithm for
factoring and Simon's algorithm for the hidden XOR-secret problem.

\subsection{Shor's Algorithm for Factoring}
Standard number theory reduces factoring $N$ to finding the order of a
random element $a$ modulo $N$, i.e., $r>0$ such that 
$a^r\equiv 1\;(\mod N)$ but $a^s\not\equiv1\;(\mod N)$ for any
$0<s<r$. Suppose $2^{n-1}<N\le2^n$. Shor's algorithm uses $2n$ qubits,
separated into two $n$-qubit registers. Initially the state is
initialized to $\ket{\phi_0} = \ket{0^n}\ket{0^n}$. By applying
the Fourier transformation followed by modular exponentiation, this
state is converted to 
\begin{math}
\ket{\phi_1} =
{1\over2^{n/2}}\sum_{x}\ket{x}\ket{a^x\;\mod N}.
\end{math}
Then one measures the second register and discard it, leading to a state
\begin{math}
\ket{\phi_2} = \sum_{t}\ket{t\cdot r+c}
\end{math}
for some random $c\in[r]$, where $t$ ranges from $0$ to
$\lfloor(2^n-1-c)/r\rfloor$ (we ignore the scalar
factor). Finally, one applies the inverse
Fourier transform to the first register followed by a measurement. The
distribution of the measurement result is approximately uniform over 
$\{[t\cdot 2^n/r]\::\:0\le t\le \lfloor(2^n-1-c)/r\rfloor\}$. One can
then solve $r$ from one instance of 
$[t\cdot2^n/r]$ using continued fraction.

\remove{
We show how Shor's algorithm fits into the QSCE paradigm. The input is
a pair $(N,a)$, and the ``hard function'' is the order of $a$ modulo
$N$:  $\phi(N,a)=r$. The hidden subset is
$S_{(N,a)}=\{[t\cdot 2^n/r]\::\:0\le t\le\lfloor(2^n-1-c)/r\rfloor\}$.
The quantum sampling circuit $Q_{(N,a)}$ produces a random element
that is approximately uniformly distributed in $S_{(N,a)}$. The
classical extracting circuit takes one sample from $S_{(N,a)}$ and
perform continued fraction to find $r$.
}

\subsection{Simon's Problem and Algorithm}
A function $f:\bit^n\mapsto\bit^n$ is given as an oracle, with the
promise that there exists an $s\in\bit^n$ (known as the ``hidden
secret'') 
such that  $f(x)=f(y)$ iff $x\oplus y=s$. Notice that if $s=0^n$, then
$f$ is a permutation, and otherwise $f$ is a 2-to-1 function. The
problem is to tell if $s=0^n$. 

Simon's algorithm works as follows. One
starts with $2n$ qubits, separated into two $n$-qubit registers. 
Originally one initializes the state to 
$\ket{\phi_0}=\ket{0^n}\ket{0^n}$. Next, one applies the Hadamard
operator to the first register and then the oracle operator 
$\ket{x}\ket{y}\mapsto\ket{x}\ket{f(x)\oplus y}$. 
The state becomes 
$\ket{\phi_1}={1\over2^{n/2}}\sum_{x}\ket{x}\ket{f(x)}$.
Next, the second register is measured and
discarded. If $s=0^n$, 
then the measurement result is $\ket{\phi_2}=\ket{x}$ for a
random $x\in\bit^n$. If $s\ne 0^n$, then the measurement is 
\begin{math}
\ket{\phi_2'}={1\over\sqrt{2}}(\ket{x} + \ket{x\oplus s})
\end{math} for 
a random $x$. Next, a Hadamard operator is applied to the first
register. In the case $s=0^n$, the result is $\ket{\phi_3} = \ket{y}$
for a random $y$; in the case $s\ne 0^n$, the result is 
$\ket{\phi_3'} = \ket{y}$ for a random $y$ such
that $y\cdot s=0$. Finally one measures the first register and obtains
$y$. Repeating the experiment $O(n)$ times,
one can solve for $s$ by using Gaussian elimination and distinguish
the case $s=0^n$ from the case $s\ne0^n$.

\remove{
We show how Simon's algorithm fits into the QSCE paradigm. The input
is given as an oracle $\oracle_f$ which computes maps $\ket{x}\ket{y}$
to $\ket{x}\ket{f(x)\oplus y}$, where $f$ has a hidden parameter
$s$. The 
``hard function'' $\phi(\oracle_f)$ evaluates to $0$ if the hidden
parameter $s$ is $0^n$, and to $1$ otherwise. The hidden subset is
$S_{\oracle_f} =\{y\in\bit^n\backslash\{0^n\}\:|\:y\cdot s=0\}$. The
quantum sampling circuit $Q_{\oracle_f}$ produces random samples
uniformly distributed in 
$S_{\oracle_f}$. The classical extracting circuit takes $O(n)$ samples
produced by $Q_{\oracle_f}$, performs the Gaussian elimination, solves
$s$, and computes $\phi(\oracle_f)$.
}

\section{The Hoeffding Bound}
\label{app:hoeffding}

We state the Hoeffding Bound, a classical result in estimating
tail probabilities.

\begin{lemma}[Hoeffding Bound~\cite{Hoe63}]
\label{lemma:hoeffding-1}
Let $k = (p-\epsilon)n$, where $\epsilon$ is a real number between
$0$ and $1/2$, and $p$ is a real number between 0 and 1. We have
\begin{equation}
\sum_{j=0}^{k}{n\choose j}p^j(1-p)^{m-j}\le e^{-2n\epsilon^2}
\end{equation}
\qed
\end{lemma}

\section{Statistical Distance}
\label{app:sd}

We define the statistical distance and state some of its
properties. The definitions and the results are standard. A good
reference to the statistical distance is Vadhan's thesis~\cite{Vad00}.

\begin{definition}[Statistical Distance]
The \emph{statistical distance} between two probability
distributions $A$ and $B$, denoted as $\sd(A,B)$, is defined to be
\begin{equation}
\sd(A,B)={1\over2}\sum_{x}|A(x) - B(x)|
\end{equation}
where the summation is taken over the support of $A$ and $B$. If
$\sd(A,B)\le\epsilon$, we say $A$ is \emph{$\epsilon$-close} to $B$.
\end{definition}

This definition can be easily extended to the continuous case with the
summation being replaced by integral and the distributions replaced by
density functions.

\begin{lemma}
\label{lemma:sd-prob}
Let $T(x)$ be a probabilistic  event with $x$ as input. Let $A$ and
$B$ be two distributions. We have
\begin{equation}
\left| \prob_{x\in A}[T(x)] - \prob_{x\in B}[T(x)]\right|
\le \sd(A,B)
\end{equation}
\qed
\end{lemma}

\begin{lemma}[Sub-additivity]
\label{lemma:sub-add}
Let $A_1$, $A_2$, $B_1$, $B_2$ be distributions, then we have
\begin{equation}
\sd(A_1B_1, A_2B_2) \le \sd(A_1,A_2) +\sd(B_1,B_2)
\end{equation}
where $AB$ denotes the \emph{tensor product} of the distributions
$A$ and $B$, i.e., $AB(a,b) = A(a)\cdot B(b)$.
\qed
\end{lemma}

\begin{lemma}
\label{lemma:uniform-sd}
Let $D_1$ be a uniform distribution over an interval $[a,a+l]$ and
$D_2$ a uniform distributions over $[b,b+l]$. Then $\sd(D_1,D_2)$ is
at most $|a-b|/l$. 
\end{lemma}
\begin{proof}
Notice that both $D_1$ and $D_2$ are uniform distributions of same
length, and thus their density functions have value $1/l$ over their
supports and 0 elsewhere. Consider the absolute difference between the
two density functions,  $|D_1(x)-D_2(x)|$. The size of its support is
at most $2|a-b|$. Thus $\sd(D_1,D_2)\le |a-b|/l$.
\end{proof}

\section{Proofs}
\label{app:proof}

\begin{lemma}
\label{lemma:bias-fourier}
Let $\hat{\dom}_{n,p}$ be the domain defined in (\ref{eqn:dom-blin})
and $\mathcal{L}_{n,p}$ be the class of normalized booleanized linear
functions over $\hat{\dom}_{n,p}$. For any query function
$g:\hat{\dom}_{n,p}\mapsto\bit$, there are at most $p^{2n/3+2}$
predicates in $\mathcal{L}_{n,p}$ that are not $1/p^{n/3}$-independent
from $g$.
\end{lemma}

For the proof we will need:

\begin{lemma}[\textbf{\cite{Yan01}}]
\label{lemma:correlation}
Let $\Omega=\{f_i\}$ be a set of function of range $\ebit$
 and $d$ be its 
cardinality. If $\langle f_i, f_j\rangle=\lambda$ for all $i\ne j$,
then the set $\{\tilde{f}_i\}$ forms an orthonormal basis for the
linear space spanned by $\Omega$, where
\begin{equation}
\tilde{f}_i(x) = 
{1\over{\sqrt{1-\lambda}}}f_i(x) -{1\over d}\cdot
\left( {1\over{\sqrt{1-\lambda}}} - {1\over{\sqrt{1+(d-1)\lambda}}}
\right)
\cdot\sum_{j=1}^df_j(x)
\end{equation}
\qed
\end{lemma}

\begin{proofof}{Lemma \ref{lemma:bias-fourier}}
We first slightly modify the class $\mathcal{L}_{n,p}$ so that its
range becomes $\ebit$. We define $\tilde{L}_{\bvec{a}}(\bvec{x}) =
2\cdot {L}_{\bvec{a}}(\bvec{x})-1$.  It is not hard to see that each
of the $p^{n-1}$ normalized booleanized linear functions maps
a $1/p$ fraction of the elements in $\hat{\dom}_{n,p}$ to $+1$, and a
straightforward but tedious analysis (see~\cite{Yan01} for a detailed
account) shows that any two normalized booleanized linear functions
agree at exactly $(p^2-2p+2)p^{n-2}-p$
places in $\hat{\dom}_{n,p}$. 
We define an inner product between functions over
$\hat{\dom}_{n,p}$ as 
\begin{equation}\langle f,g\rangle={1\over
  p^n-p}\sum_{x\in\hat{\dom}_{n,p}}f(x)g(x),
\end{equation}
With this inner product, any query function has norm 1, and any pair
of distinct functions
$\tilde{L}_{\bvec{a}}$ and $\tilde{L}_{\bvec{b}}$ have the same
inner product.  This will allow us to 
``extract'' an orthonormal basis from the class
$\mathcal{L}_{n,p}$ using Lemma~\ref{lemma:correlation}.

Now we fix a query function $g$ and relate predicates that are not
$\xi$-independent from $g$ to the Fourier coefficients of $g$. 
Consider a booleanized linear function $L_{\bvec{a}}$, and we denote
its positive set by $S$. We have that $|S|=p^{n-1}-1$.
Suppose $g$ maps $a$ elements in $\hat{\dom}_{n,p}$ to $+1$, and $b$
elements in $S$ to $+1$. Then if $L_{\bvec{a}}$ is not
$\xi$-independent from $g$, we have
\begin{equation}
\left|{2a-p^n+p\over p^n-p}-{2b-p^{n-1}+1\over p^{n-1}-1}\right|>\xi,
\end{equation}
or $|a-bp|>{p^n-p\over2}\xi$. We write $b=a/p+\delta$, and we have
$|\delta|\ge {p^{n-1}-1\over2}\xi$.

Next we compute the inner product of $g$ and $\tilde{L}_{\bvec{a}}$.
Straightforward computation shows that
\begin{eqnarray*}
\hspace{-0.2in}
\langle g, \tilde{L}_{\bvec{a}} \rangle  
& = & 
2\cdot\left(2b-a+(p-1)(p^{n-1}-1)\over p^n-p
\right)-1 \\
& = & 
\left(1-{2a\over{p^n-p}}\right)\left(1-{2\over p}\right) +
{4\delta\over p^n-p}
\end{eqnarray*}

On the other hand, the inner product of $g$ with an \emph{average}
over booleanized linear functions is 
\begin{eqnarray*}
\hspace{-0.2in}
{1\over p^{n-1}}\sum_{\bvec{b}[0]=1}\langle g, \tilde{L}_{\bvec{b}} 
\rangle & = & {1\over p^{n-1}(p^n-p)}\sum_{\bvec{b}[0]=1}
\sum_{x\in\hat{\dom}_{n,p}}g(x) \tilde{f}_{\bvec{b}}(x)\\
& = & {1\over p^{n-1}(p^n-p)}\sum_{x\in\hat{\dom}_{n,p}}g(x)
\sum_{\bvec{b}[0]=1}\tilde{f}_{\bvec{b}}(x)\\
& = & \left(1-{2a\over p^n-p}\right)\left(1-{2\over p}\right)
\end{eqnarray*}

Now we apply Lemma~\ref{lemma:correlation}, setting $d=p^{n-1}$ and
$\lambda={(p^2-4p+4)p^{n-2}-p\over p^n-p}$. We will obtain an
orthonormal basis, which we denote by $\{\hat{L}_{\bvec{b}}\}$.

Putting things together, we can compute that Fourier coefficient of
$g$ over the component $\hat{L}_{\bvec{a}}$.
\begin{eqnarray*}
\hspace{-0.2in}
\langle g, \hat{L}_{\bvec{a}} \rangle & = & 
{1\over \sqrt{1-\lambda}}\langle g, \tilde{L}_{\bvec{a}} \rangle - 
\\ & & 
\left( {1\over{\sqrt{1-\lambda}}} -
{1\over{\sqrt{1+(d-1)\lambda}}}
\right)\cdot {1\over d}\sum_{\bvec{b}[0]=1}\langle 
g, \tilde{L}_{\bvec{b}} \rangle\\
& = & 
{1\over \sqrt{1-\lambda}}\cdot \left[
\left(1-{2\over p}\right)\cdot\left(1-{2a\over
    p^n-p}\right)+{4\delta\over p^n-p}
\right] - \\
& & 
\left( {1\over{\sqrt{1-\lambda}}} - {1\over{\sqrt{1+(d-1)\lambda}}}
\right)\cdot \left(1-{2\over p}\right)\cdot 
\left(1-{2a\over p^n-p}\right)
\\
& = & {1\over{\sqrt{1+(d-1)\lambda}}}\left(1-{2\over p}\right)\cdot
\left(1-{2a\over p^n-p}\right)
+ \\
& & {1\over \sqrt{1-\lambda}}\cdot {4\delta\over p^n-p} \\
& = & {1\over p^{(n-1)/2}}\left(1-{2a\over p^n-p}\right) + 
\\ & & 
{2\delta\over\sqrt{p}(p^{n-1}-1)}\cdot\sqrt{1-1/p^{n-1}\over1-4/p} \\
& \ge &  {2\delta\over\sqrt{p}(p^{n-1}-1)} - {1\over p^{(n-1)/2}}
\end{eqnarray*}
Now we substitute in $\xi=1/p^{n/3}$, and we have
\begin{equation}
|\langle g, \hat{L}_{\bvec{a}} \rangle| \ge {\xi\over\sqrt{p}}-
{1\over   p^{(n-1)/2}} \ge {1\over p^{n/3+1}}
\end{equation}

Thus $g$ can
have at most $p^{2n/3+2}$ such Fourier coefficients, and so there can
be at most $p^{2n/3+2}$ predicates that are not $1/p^{n/3}$-independent
from $g$.
\end{proofof}

\begin{lemma}
\label{lemma:independent-count}
Let $\dom_n = \bit^n\backslash\{0^n\}$ and $\funcc_n$ be the class of
negative parity functions over $\dom_n$. For any query function
$g:\bit^n\mapsto\ebit$, there are at most $2^{n/2+2}$ predicates in
$\funcc_n$ that are not $2^{-n/4}$-independent from $g$.
\end{lemma}
\begin{proof}
We fix a negative parity function $f$.
Let $a$ denote the number of $x\in\bit^n$ such that $g(x)=1$, and let
$b$ denote the number of $x\in S_f$ such that $g(x)=1$.  Notice that
since all parity functions are balanced, we have $|S_f|=2^{n-1}-1$
(since $f(0^n)=1$ but $0^n\not\in S_f$). Then if $f$ is not
$\xi$-independent from $g$, we have
\begin{equation}
\label{eqn:independent}
\left|{2b-2^{n-1}+1\over2^{n-1}-1} - {2a - 2^n\over
    2^n}\right|>\xi
\end{equation}
or
\begin{equation}
\label{eqn:independent-2}
\left|{a-2b\over 2^{n-1}-1}\right|> \xi - {a\over2^{n-1}(2^{n-1}-1)}
> \xi- {1\over{2^{n-1}-1}}
\end{equation}

Next we perform Fourier analysis. We first define an inner product of
real functions over $\bit^n$: 
\begin{equation}
\langle f,g\rangle={1\over2^n}\sum_{x\in\bit^n}f(x)g(x).
\end{equation}
We define a set of ``modified parity functions'' as
$\tilde{\oplus}_s(x) = (-1)^{s\cdot x}$, which map elements in
$\bit^n$ to $\ebit$. It is clear that the set of all parity functions
$\{\tilde{\oplus}_s(x)\}_s$ form an orthonormal basis, and
$\tilde{\oplus}_s(x)=1-2\lnot\oplus_s(x)$. If a parity function
$\lnot\oplus_s(x)$ is not $\xi$-independent from $g$, then
(\ref{eqn:independent}) holds (by setting $f=\lnot\oplus_s$).
Let $t=g(0^n)$. Within the subset where $\tilde{\oplus}_s(x)=-1$,
which includes $0^n$ and the positive set of $\lnot\oplus_s$, $g$ maps
$b+t$ inputs to $+1$. Outside this subset, $g$ maps $a-b-t$ inputs to
$+1$, and $2^{n-1}-a+b+t$ input to $-1$. Thus, we can compute the
Fourier coefficient of $g$ on $\tilde{\oplus}_s$.
\begin{eqnarray*}
\langle \tilde{\oplus}_s, g\rangle  & =  & 
1-2\cdot \prob_{x\in\bit^n}[\tilde{\oplus}_s(x) = g(x)] \\
& = & 1-2\cdot\left(
{a-b-t\over2^n} + {2^{n-1}-a+b+t\over 2^n}\right)\\
&  = & {2a-4b-4t\over2^n}
\end{eqnarray*}
Substituting in (\ref{eqn:independent-2}), we have 
\begin{equation}
\label{eqn:fourier-coef}
|\langle \tilde{\oplus}_s, g\rangle|> \xi - {6/2^n}.
\end{equation}
However, notice that the query function $g(x)$ has norm 1 and thus it
can have at most $1/(\xi - {6/2^n})^2$ Fourier coefficients such that
(\ref{eqn:fourier-coef}) holds. Now plugging in $\xi = 2^{-n/4}$, we
have
$1/(\xi - {6/2^n})^2\le 2^{n/2+2}$, and the Lemma is proved.

\end{proof}

\end{document}